\documentclass[prl,twocolumn,showpacs,amsmath,amssymb,superscriptaddress]{revtex4}
\usepackage{graphicx}% Include figure files
\usepackage{dcolumn}% Align table columns on decimal point
\usepackage[sort&compress]{natbib}
\usepackage{subfigure}
\usepackage{ifpdf}
\usepackage{bm}
\begin{document}
\title{The average kinetic energy density of Cooper pairs above $T_c$ in ${\rm YBa_2Cu_3O_{7-x}}$,
${\rm Bi_2Sr_2CaCu_2O_{8+\delta}}$, and ${\rm Nb}$}
\author{S. Salem-Sugui Jr.}
\affiliation{Instituto de F\' \i sica, Universidade Federal do Rio
de Janeiro, 21941-972, Rio de Janeiro, RJ, Brazil}
\author{Mauro M. Doria}
\affiliation{Instituto de F\' \i sica, Universidade Federal do Rio
de Janeiro, 21941-972, Rio de Janeiro, RJ, Brazil}
\affiliation{Departement Fysica, Universiteit Antwerpen,
Groenenborgerlaan 171, B-2020 Antwerpen, Belgium}
\author{A. D. Alvarenga}
\affiliation{Instituto Nacional de Metrologia Normaliza\c{c}\~ao e
Qualidade Industrial, Duque de Caxias, 25250-020, RJ, Brazil.}
\author{V. M. Vieira}
\affiliation{ Departamento de F\' \i sica, Universidade Federal de
Pelotas, RS, Brazil}
\author{P. F. Farinas}
\affiliation{Instituto de F\' \i sica, Universidade Federal do Rio
de Janeiro, 21941-972, Rio de Janeiro, RJ, Brazil}
\author{J. P. Sinnecker}
\affiliation{Instituto de F\' \i sica, Universidade Federal do Rio
de Janeiro, 21941-972, Rio de Janeiro, RJ, Brazil}
\date{\today}
\begin{abstract}

We have obtained isofield curves for the square root of the average
kinetic energy density of the superconducting state for three single
crystals of underdoped $YBa_2Cu_3O_{7-x}$, an optimally doped single
crystal of $Bi_2Sr_2CaCu_2O_{8+\delta}$, and Nb. These curves,
determined from isofield magnetization versus temperature
measurements and the virial theorem of superconductivity, probe the
order parameter amplitude near the upper critical field. The
striking differences between the Nb and the high-$T_c$ curves
clearly indicate for the latter cases the presence of a unique
superconducting condensate below and above $T_c$.
\end{abstract}
\pacs{{74.25.Bt},{74.25.Ha},{74.72.Bk},{74.62.-c}} \maketitle

A considerable amount of evidence points towards the existence of
superconductivity above the superconducting transition, $T_c$.
Nernst coefficient measurements done in different compounds
\cite{wang1,pourret}, and also torque magnetometry in
$Bi_2Sr_2CaCu_2O_{8+\delta}$ (Bi 2212) \cite{wang1} indicate a state
with non-zero amplitude and phase incoherence \cite{wang1,wang2},
suggestive of a three-dimensional Kosterlitz-Thouless scenario
inside the pseudo-gap region \cite{lee,wang1,timusk}. According to
the BCS theory the electronic state acquires kinetic energy
\cite{corson} upon condensation, and for the high-$T_c$ compounds
infrared and reflectivity measurements show a large transfer of
spectral weight to the superfluid condensate
\cite{corson,santander}, supportive of an in-plane kinetic energy
driven mechanism. This kinetic energy gain holds for the overdoped
Bi2212 compound but not for the optimally doped and the underdoped
compounds \cite{deutscher}, where is in fact a loss. Thus the
kinetic energy is a relevant tool to sort the distinct proposals for
the normal state and their consequent pairing mechanisms
\cite{deutscher, santander}. Despite that, the literature lacks
experimental information about the kinetic energy of the condensate
in presence of an applied field, $K$, although a method to determine
it has been proposed a few years ago \cite{mauro}. In this letter we
show that this method is a tool to obtain information about the
condensate above $T_c$. For instance, $K$ provides information on
the amplitude of the order parameter near $T_c$, a quantity of first
importance to understand the nature of the superconducting state
below and above $T_c$. There are several reports of a pseudo-gap
that extends superconductivity above $T_c$ for the high-$T_c$
compounds \cite{rotter,ding,timusk,renderia}, and they basically
split into two views \cite{pines}, either as directly related to the
superconducting state or as a competing independent effect. The
present results clearly support the first view although we do not
directly probe the pseudogap in our study.

In this letter we obtain isofield curves of $\sqrt K$ vs. $T$ and
show that this quantity smoothly evolves from below to above $T_c$,
without any abrupt change, even in its first derivative with respect
to the temperature, strongly suggestive of the existence of a unique
condensate state below and above $T_c$ for $YBa_2Cu_3O_{7-x}$
(YBaCuO) and for Bi2212. For comparison we also study a low-$T_c$
superconductor, e.g., Niobium (Nb), and find that our procedure is
able to reproduce the standard BCS behavior. We chose to study
$\sqrt K$ because of its direct relation to the amplitude of the
order parameter near $T_c$. For zero field $\sqrt K$ proportional to
the superconducting energy gap, $\Delta(T)$, through the BCS
expression $K \sim \Delta^2$ \cite{degennes}.
%/////////////////fig Nb ///////////////////////////////////
\begin{figure}[t]
% Requires \usepackage{graphicx}
\includegraphics[width=0.95\linewidth]{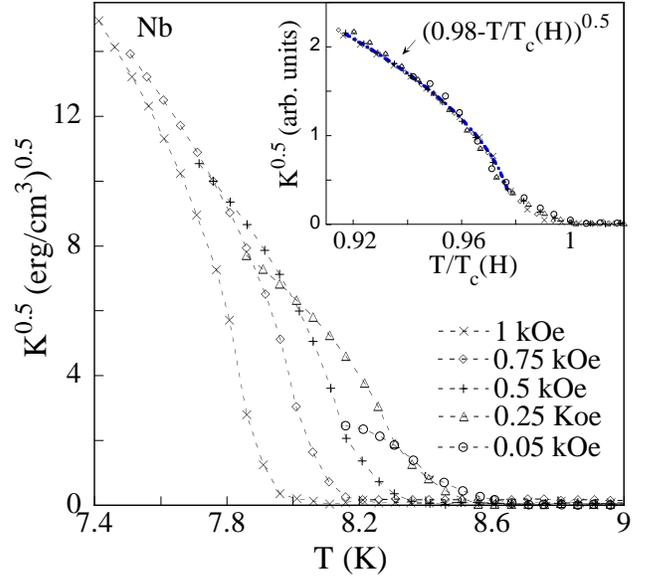}
\caption{Selected isofield $\sqrt K$ versus temperature curves are
shown for Nb ($T_c= 8.5 K$). The inset displays the collapse of
these curves upon scaling and shows the fit agreement with the
BCS-Abrikosov theory.} \label{nb}
\end{figure}
%%%%%%%%%%%%%%%%%%%%%%%%%%%%%%%%%%%%%%%%%%%%%%%%%%%%%%%

Average quantities can be determined in many-body physics by the
virial theorem even in cases that knowledge about the interaction
among particles is not complete. In classical systems the virial
theorem leads to remarkable estimates, e.g., the interior
temperature of the Sun \cite{berkeley}, and an upper bound to the
mass of a white dwarf star - the Chandrasekhar limit\cite{collins}.
Interestingly, the average kinetic energy of the condensate can be
directly retrieved from the equilibrium magnetization, $M$, for a
large $\kappa$ type II superconductor, in the pinning free
(reversible) regime \cite{mauro}. This connection is a direct
consequence of the virial theorem \cite{doria1}:
\begin{equation}
K = <\frac{\hbar^2}{2m}|({\bm \nabla}-\frac{2\pi i}{\Phi_0}{\bm
A})\psi|^2 > = B.|M|,\label{kin1}
\end{equation}
where $\psi$ is the order parameter and $B$ is the magnetic
induction.
%/////////////////fig mvst ///////////////////////////////////
\begin{figure}[t]
% Requires \usepackage{graphicx}
\includegraphics[width=\linewidth]{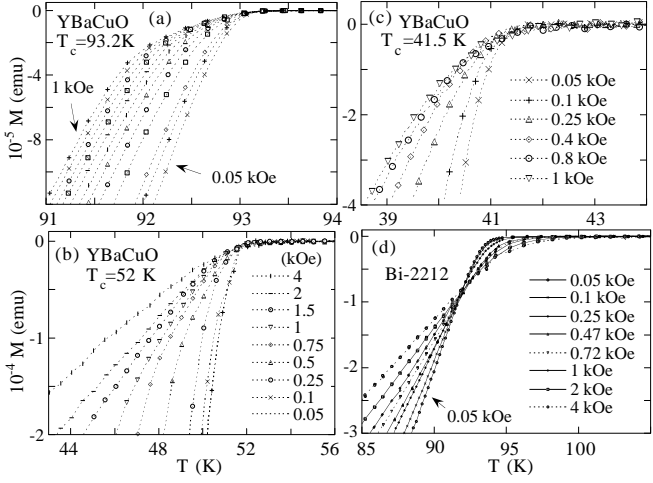}
\caption{Magnetization versus temperature are shown for several
applied fields for YBaCuO and  Bi2212 single crystals.} \label{mvst}
\end{figure}
%%%%%%%%%%%%%%%%%%%%%%%%%%%%%%%%%%%%%%%%%%%%%%%%%%%%%%%%
The Abrikosov treatment \cite{abrikosov} of the Ginzburg-Landau
theory gives that the average kinetic energy density of the
condensate near to the transition is,
\begin{equation}
K = \frac{p^2}{2m}
<|\psi|^2>=\frac{H[H_{c2}(T)-H]}{8\pi\kappa^2\beta_A},\label{kin2}
\end{equation}
where $<\cdots>$ means spatial average, $p=\hbar\sqrt{2\pi
H/\Phi_0}$, $H_{c2}(T)=\Phi_0/2\pi\xi(T)^2$, and $\beta_A \simeq 1$
is the lattice constant. Thus in this case too, $\sqrt K$ is
proportional to the average order parameter amplitude, and so, to
the average superconducting energy gap. The general virial relation
(Eq. \ref{kin1}) applies throughout the mixed state and,
consequently, reduces to Eq. \ref{kin2} near the upper critical
field, $H_{c2}(T)$. We find that the isofield $\sqrt K$ vs. $T$
curves are well fitted by Eq. \ref{kin2} for Nb with the expected
coherence length temperature behavior ($\xi(T) \sim
(T-T_c)^{-0.5}$), thus confirming a BCS gap behavior. However this
is not the case for the high-$T_c$ compounds, whose  average kinetic
energy density  does not vanish at the BCS temperature $T_c(H) <
T_c(0)=T_c$, but at some higher temperature, $T^e$, above $T_c$.
Therefore we show here that Eq. \ref{kin1} is able to unveil the
unusual properties of the gap in presence of an applied field for
the high-$T_c$ compounds.
%/////////////////fig y41 ///////////////////////////////////
\begin{figure}[t]
% Requires \usepackage{graphicx}
\includegraphics[width=\linewidth]{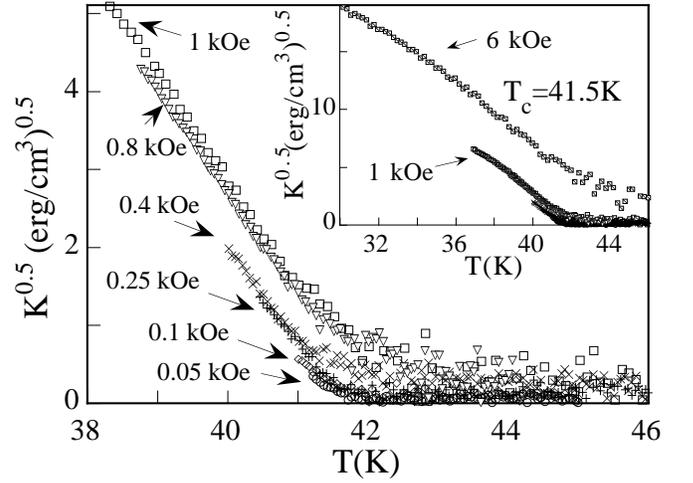}
\caption{Isofield $\sqrt K$ versus temperature curves are shown for
the YBaCuO compound. The inset displays an extra curve taken at a
field above those shown in the main figure.} \label{y41}
\end{figure}
%/////////////////fig y52 ///////////////////////////////////
\begin{figure}[t]
% Requires \usepackage{graphicx}
\includegraphics[width=\linewidth]{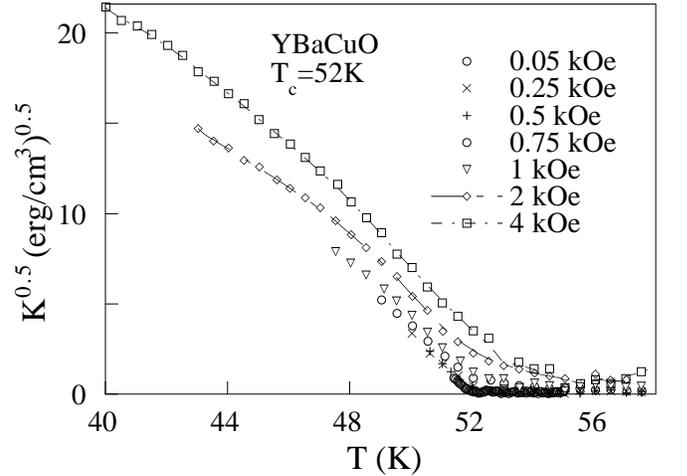}
\caption{Isofield $\sqrt K$ versus temperature curves as obtained
for the YBaCuO compound } \label{y52}
\end{figure}
%/////////////////fig y93 ///////////////////////////////////
\begin{figure}[t]
% Requires \usepackage{graphicx}
\includegraphics[width=\linewidth]{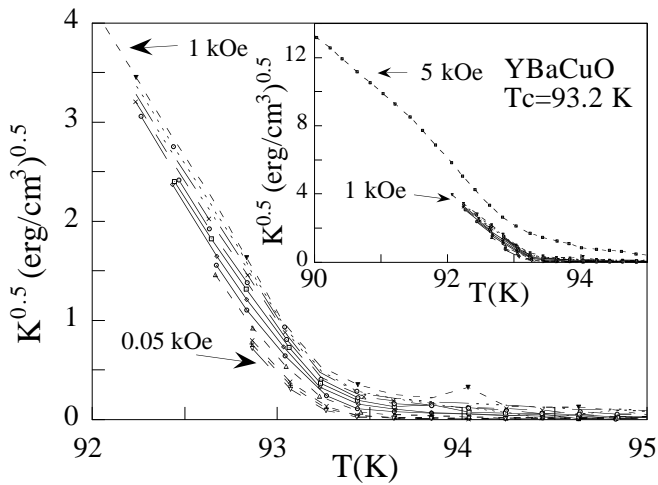}
\caption{Isofield $\sqrt K$ versus temperature curves are shown for
the YBaCuO compound. The inset displays an extra curve taken at a
field above those shown in the main figure.} \label{y93}
\end{figure}
%%%%%%%%%%%%%%%%%%%%%%%%%%%%%%%%%%%%%%%%%%%%%%%%%%%%%%%

We measured three single crystals of $YBa_{2}Cu_{3}O_{7-x}$ ($T_c$=
93.2 K for $x\sim 0.05$; $T_c$= 52 K  for $x=0.5$, and $T_c$=41.5 K
for $x=0.6$), a single crystal of Bi2212 ($T_c$=93 K) and a Nb
sample ($T_c$=8.5 K and Ginzburg-Landau parameter $\kappa=4$)
\cite{mauro}. The single crystals of YBaCuO and Bi2212 were grown at
Argonne National Laboratory \cite{veal} and exhibit fully developed
transitions with ($\Delta T_{c}\simeq 1-2K$) for YBaCuO samples and
($\Delta T_{c}\simeq 5K$) for Bi2212. The value of $T_c$ for each
sample is determined as the onset temperature of diamagnetism, above
which magnetization values falls within the equipment sensitivity.
Measurements of the isofield $M$ vs. $T$ curves were done on (MPMS
Quantum Design and Lake-Shore) commercial magnetometers based on the
superconducting quantum interference device (SQUID). Experiments
were conducted for magnetic field values ranging from 0.05 kOe to
6.0 kOe, always applied along the $c$ axis direction of the
crystals. Magnetization data, $M$ vs. $T$  or $M$ vs. $H$ curves,
was always taken after cooling the sample below $T_{c}$ in zero
magnetic field (ZFC). To determine the reversible (equilibrium)
magnetization temperature range, we have also field-cooled the
samples from above to below $T_{c}$. The non-superconducting
background was removed for each $M$ vs. $T$ curve, by fitting the
magnetization in a temperature range well above $T_c$ ($\sim10K$ for
YBaCuO and $\sim20 K$ for Bi2212) to $M_{back}=A(H)$, where
$A(H)=a-bH$ is a constant value for each field and $a$ and $b$ are
constants determined for each sample. Above $H=1kOe$ an additional
term $C(H)/T$ had to be considered for deoxigenated YBaCuO, with
$C(H)$ very small \cite{said}. The determination of the average
kinetic energy density requires that a few low temperature $M$ vs.
$H$ curves be measured in the Meissner region as a function of the
applied field. These curves were obtained for all samples and
yielded the geometric factors $d \equiv -H/M$ with a good
resolution. Next the geometrical factor is used to obtain the
magnetic induction, $B\equiv H+d.M$, which, by its turn, is used to
determine the $\sqrt{K}$ vs. $T$ curves.
%/////////////////fig bi ///////////////////////////////////
\begin{figure}[t]
% Requires \usepackage{graphicx}
\includegraphics[width=\linewidth]{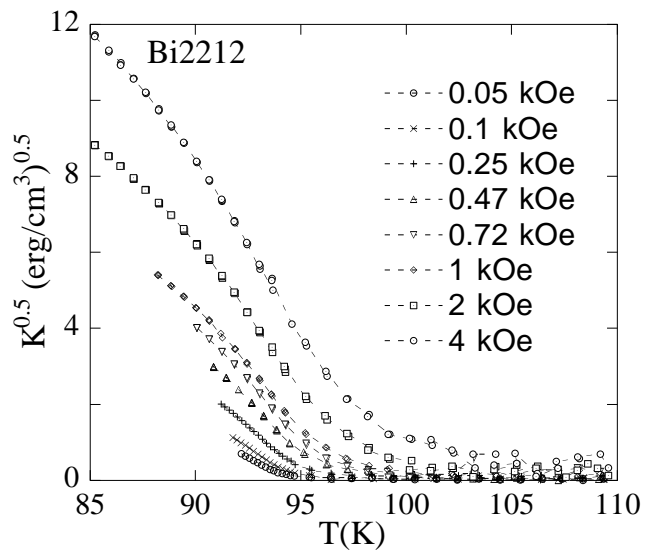}
\caption{Isofield $\sqrt K$ versus temperature curves as obtained
for the Bi2212 compound.} \label{bi}
\end{figure}
%%%%%%%%%%%%%%%%%%%%%%%%%%%%%%%%%%%%%%%%%%%%%%%%%%%%%%%%%%%%%%%%%%%%%%%%%
Our Nb study is summarized in Fig. \ref{nb}, which shows selected
isofield $\sqrt K$ vs. $T$ curves obtained from standard $M$ vs. $T$
curves. We observe that the extrapolation of the isofield $\sqrt K$
curves to the X axis ($T$ axis) matches the $T_{c}(H)$ values
obtained from the standard extrapolation of the magnetization curves
to zero. Notice the collapse of all experimental different curves
into a single one by a unique prescription, namely, by dividing the
Y axis of each curve by a fixed number and re-scaling T as
$T/T_{c}(H)$. The resulting collapsed curve for Nb is shown in the
inset of Fig. \ref{nb} and indeed shows the behavior $\sqrt K \sim
(T-T_c)^{0.5}$ for $T$ close to $T_c$, confirming that this quantity
is proportional to the gap $\Delta(T)$ \cite{degennes}. Hence it
carries information on the amplitude of the order parameter in the
vicinity of $T_c$ even for intermediated values of the applied
magnetic field.
%%%%%%%%%%%%%%%%%%%%%%%%%%%%%%%%%%%%%%%%%%%%%%%%%%%%%%%%%%%%%

The $\sqrt K$ curve leads to the average superconducting gap in
presence of an applied field, according to the BCS-Abrikosov theory.
This assertion has far reaching consequences for the high $T_c$
superconductors, but we do not directly rely on it to do our data
analysis, which is fully based on the general virial relation (Eq.
\ref{kin1}). Fig. \ref{mvst} displays the $M$ vs. $T$ reversible
curves obtained for YBaCuO and Bi2212 samples. Our measurements
reproduce standard and well known features below $T_c$, such as the
nearly field independent crossing point \cite{bulaevskii,tesanovic}
for Bi2212. For YBCO \cite{said1,rosenstein} we also observe it for
fields above 1.0 kOe. These crossing points are caused by thermal
fluctuations on vortices and will not be discussed here. We reach
our main results in Figs. \ref{y41}, \ref{y52}, \ref{y93} and
\ref{bi}, which show the $\sqrt K$ vs. $T$ curves for the three
doped YBaCuO and the Bi2212 compounds, respectively. They were
obtained from the corresponding $M$ vs. $T$ curves displayed in Fig.
\ref{mvst}. The $d\sqrt{K}/dT<0$ feature is common to all samples
including Nb. However a remarkable and striking difference exists
between the high-$T_c$ compounds and Nb. \emph{The temperature that
$\sqrt K$ extrapolates to zero, which indicates $T_{c}(H)$ for Nb
and should decrease with $H$, instead, increases with $H$ for the
YBCO and Bi2212 compounds!}.  The $\sqrt K$ vs. $T$ YBaCuO and
Bi2212 isofield curves enter the region above $T_c$ smoothly,
without any slope change. Then, owing to the present method, the
state observed above $T_c$ is found to be a continuous evolution of
the condensate state below $T_c$. We find here that the isofield
curves form a set of non-intersecting lines with roughly the same
slope that eventually disappear inside a common background that
surrounds the $T$ axis. An increase in the applied field causes
their shift to an upper region of the $\sqrt K$ vs. $T$ diagram. The
extrapolated intercept of an isofield curve with the $T$ axis
defines a new temperature $T^e$, always found to be higher than
$T_c$. Consider a $T>T_c$ vertical line in the $\sqrt{K}$ vs. $T$
diagram. The intersect of the $\sqrt K$ curves with this line
defines a function $\sqrt{K}(H)$ that grows monotonically with $H$,
in qualitative agreement with the $\sqrt{K} \sim \sqrt{H}$
prediction of Eq. \ref{kin2}. For temperatures above $T^e$ the
background noise takes over the magnetization signal rendering
impossible any further analysis near to the $T$ axis. The
non-intersecting feature of the $\sqrt K$ vs. $T$ curves is a
noticeable fact because it leads to the concept of the threshold
field needed to produce a $\sqrt K$ vs. $T$ curve above a given
temperature $T^e$. For instance, the extrapolation of the Bi2212
curve H=4.0 kOe, shown in Fig. \ref{bi}, towards zero occurs at $T^e
\approx 99.3 K$. A $\sqrt K$ curve that extrapolates to a higher
temperature, $T'^e>T^e$, must be associated to a magnetic field
larger than the threshold of 4.0 kOe.

%%%%%%%%%%%%%%%%%%%%%%%%%%%%%%%%%%%%%%%%%%%%%%%%%%%%%%%%%%%%%%%%%%
It has been long known that critical fluctuations in the vicinity of
$T_c$ can produce effects on the normal state susceptibility for
$T>T_c$ \cite{tinkham,li}, but they cannot explain our results
because the Ginzburg criterion estimates a very small temperature
window in the vicinity of $T_c$, where they are important: $G
\sim\Delta T/T_c\simeq 10^{-3}$ \cite{klemm,wlee}. Even
overestimating them, $\Delta T \simeq 1K$, keeps this window much
below the 6 K shift above $T_c$ found here for the H= 4.0 kOe Bi2212
curve, for instance.
%%%%%%%%%%%%%%%%%%%%%%%%%%%%%%%%%%%%%%%%%%%%%%%%%%%%%%%%%%%%%

Theoretically, a pseudo-gap regime is a property of 2D systems
\cite{vilk,norman}, and it is worth mentioning that a previous work
\cite{said}, performed on the same deoxygenated YBaCuO samples
($T_c$=41.5 and 52 K) used here, shows the existence of quasi-two
dimensional critical fluctuations in the region of $T_c$ for low
fields. We speculate that anisotropy plays an important role in the
shift of the isofield $\sqrt K$ vs. $T$ curves above $T_c$ because
doping effects on the pseudo-gap \cite{norman} are associated to
dimensionality. Indeed a direct inspection of Figs. \ref{y41},
\ref{y52} and \ref{y93} for YBCO, and of Fig. \ref{bi} for Bi2212,
show a correlation between the anisotropy ratio parameter, $\gamma$,
and the relative temperature deviation above the critical
temperature, $(T^e-T_c)$, of the $\sqrt K$ vs. $T$ curves. These
figures show that the Bi2212 ($\gamma \sim 200$ \cite{Iye}) has the
most accentuated deviation, with $(T^e-T_c)\cong 6K$ for $H$=4.0
kOe, followed by the YBCO compounds in descending order of deviation
\cite{chien}: $T_c=41.5 K$ ($\gamma \sim 100$ and $(T^e-T_c)\cong
3.5 K$ for $H$= 6.0 kOe), $T_c=52 K$ ($\gamma \sim 65$ and
$(T^e-T_c)\cong 2 K$ for $H$ =4.0 kOe), and $T_c= 93.2 K$ ($\gamma
\sim 8-9 K$  and $(T^e-T_c)\cong 0.3 K$ for $H$ =5.0 kOe). We stress
that at $T_c$, $\sqrt K$ grows continuously with field, and when the
maximum is reached, this is the highest used field. Important, the
later clearly indicates a finite order parameter amplitude at $T_c$,
and consequently, an upper critical field $H_{c2}$ at $T_c$ larger
at least than the largest fields used here. Thus our low field
results are supportive of the high field scenario developed by Wang
et al. \cite{wang1,wang2}.

%%%%%%%%%%%%%%%%%%%%%%%%%%%%%%%%%%%%%%%%%%%%%%%%%%%%%%%%%%%%%%%%%%
In summary, we have shown that the average kinetic energy density of
the condensate exists above the critical temperature for the
high-$T_c$ but not for the low-$T_c$(Nb) compounds. It gives further
evidence of a pseudo-gap, which has been directly measured by others
\cite{rotter,ding,timusk} in similar samples.
%%%%%%%%%%%%%%%%%%%%%%%%%%%%%%%%%%%%%%%%%%%%%
\begin{acknowledgments}
We thank Boyd Veal and D. G. Hinks from Argonne National Laboratory,
who kindly provide the high-$T_c$ single crytals. We thank CNPq,
FAPERJ  and the Instituto do Mil\^enio de Nanotecnologia for
financial support.
\end{acknowledgments}

\end{document}